\documentclass[aps,prb,twocolumn,letterpaper,superscriptaddress]{revtex4}

\usepackage{times,xspace}
\usepackage{amsbsy,amssymb,bm,mathptmx}
\usepackage{graphicx,color,epsfig,rotate}
\usepackage{fancyhdr}
\usepackage{amsmath,amssymb,amsfonts,amsthm}
\usepackage{tabularx,siunitx}
\usepackage{booktabs}
\renewcommand{\vec}[1]{\bm{#1}}

\begin{document}

\title{The macroscopic monopolization in diagonal magnetoelectrics}

\date{\today}

\author{Nicola A.~Spaldin}
\affiliation{Materials Theory, ETH Zurich, Wolfgang-Pauli-Strasse 27, 8093 Zurich, Switzerland}
\author{Michael Fechner}
\affiliation{Materials Theory, ETH Zurich, Wolfgang-Pauli-Strasse 27, 8093 Zurich, Switzerland}
\author{Eric Bousquet}
\affiliation{Materials Theory, ETH Zurich, Wolfgang-Pauli-Strasse 27, 8093 Zurich, Switzerland}
\affiliation{Physique Th\'eorique des Mat\'eriaux, Universit\'e de Li\`ege, B-4000 Sart Tilman, Belgium}
\author{Alexander Balatsky}
\affiliation{NORDITA, KTH Royal Institute of Technology and Stockholm University, Roslagstullsbacken 23 106 91 Stockholm, Sweden}
\affiliation{Theoretical Division, Los Alamos National Laboratory, Los Alamos,
NM, 87545, USA}
\affiliation{Center for Integrated Nanotechnologies, Los Alamos National Laboratory,
Los Alamos, NM 87545, USA}
\author{Lars Nordstr\"om}
\affiliation{Department of Physics and Astronomy, Uppsala University, P.O. Box 516, SE-75120 Uppsala, Sweden}
 
\begin{abstract}
We develop the formalism of the macroscopic monopolization -- that is the monopole moment per unit volume -- in 
periodic solids, and discuss its relationship to the diagonal magnetoelectric effect. For the series of lithium 
transition metal phosphate compounds we use 
first-principles density functional theory to calculate the contributions to the macroscopic monopolization
from the global distribution of magnetic moments within the unit cell, as well as from the distribution of
magnetization around the atomic sites. We find one example within the series (LiMnPO$_4$) that
shows a macroscopic monopolization corresponding to a ferromonopolar ordering consistent with its diagonal 
magnetoelectric response. The other members
of the series (LiMPO$_4$, with M = Co, Fe and Ni) have zero net monopolization but have antiferromonopolar
orderings that should lead to $q$-dependent diagonal magnetoelectric effects.
\end{abstract}

\pacs{}

\maketitle

\section{Introduction}
\label{intro}

The linear magnetoelectric response of a solid is the linear order magnetization 
induced by an electric field or equivalently the linear order electric polarization 
induced by a magnetic field. It is described by a second-rank tensor, $\alpha$, which 
can be non-zero when both time-reversal and space-inversion symmetries are broken, and 
may have diagonal or off-diagonal components, corresponding to a response parallel or 
perpendicular to the applied field respectively. 

Materials with anti-symmetric off-diagonal linear magnetoelectric responses
have the same symmetry as the toroidal component of the second-order term
in the magnetic multipole expansion, and so there has been much recent discussion in the 
literature of whether the toroidal moment, $t$, is a relevant and useful
concept for describing such magnetoelectric effects. 
In particular, the term {\it ferrotoroidics} has been introduced to
describe materials in which the toroidal moments are aligned cooperatively,
and such materials have been considered to complete the group of primary
ferroics.\cite{Schmid:2003,Schmid:2004,VanAken_et_al:2007} Motivated by
this suggestion, a theory of {\it toroidization} -- defined to be the toroidal moment
per unit volume -- in bulk crystalline solids has
been developed, which appropriately treats the multi-valuedness caused
by the periodic boundary conditions \cite{Ederer/Spaldin:2007}. Ferrotoroidic
switching has been reported \cite{VanAken_et_al:2007}, and attempts to
demonstrate that the toroidal moment can act as a primary order parameter
are ongoing. In addition, the {\it local} toroidal moments associated with the 
atomic V sites in V$_2$O$_3$ and the
atomic Cu sites in CuO have been 
detected directly using resonant x-ray diffraction
\cite{Lovesey_et_al:2007,Fernandez-Rodriguez_et_al:2010,Scagnoli_et_al:2011}.
Such local toroidal moments could be of tremendous importance, as it has been 
proposed that they are candidates for the order parameter in the pseudo-gap 
phase of cuprate superconductors\cite{Shekhter/Varma:2009}. 

The second-order term in the magnetic multipole expansion contains two
additional contributions beyond the toroidal term, which describe in 
turn magnetic quadrupolar and magnetic monopolar components that couple
respectively to the gradient and divergence of the magnetic field (see 
detailed derivation below). While the latter has not been extensively 
discussed on the grounds that Maxwell's equations tell us formally that 
$\vec{B}$ does not diverge, it is in fact non-zero in materials
with a {\it diagonal} linear magnetoelectric response. Indeed, it could
appropriately be described as a {\it magnetoelectric monopole} to 
distinguish it from the zeroth order term in the multipole expansion
of the magnetic field which is the magnetic analogue to the electrical
charge and indeed is formally zero. We emphasize also that the magnetoelectric 
monopole discussed here is a {\it ground state property} of the
system, and so is distinct from those recently proposed and verified in spin 
ice, in which nonlocal magnetic monopoles exist as {\it excited states}
\cite{Castelnovo/Moessner/Sondhi:2008,Morris_et_al:2009}.  

\begin{figure}
\centerline{\includegraphics[width=0.95\columnwidth]{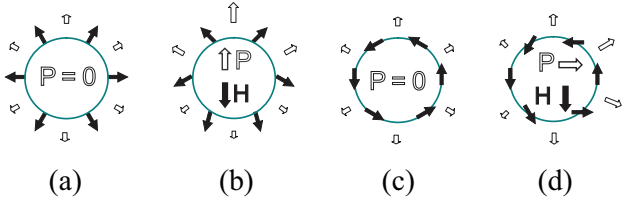}}
\caption{Diagonal ((a) and (b)) and off-diagonal ((c) and (d)) magnetoelectric responses of monopolar
and toroidal spin arrangements. From Ref. \protect \onlinecite{Delaney/Mostovoy/Spaldin:2009}.
}
\label{vortices}
\end{figure}

The origin of the relationship between the monopolar contribution
to the multipole expansion and the diagonal magnetoelectric response
is illustrated in Fig.~\ref{vortices} (a) and (b) where we follow the discussion from 
Ref.~\onlinecite{Delaney/Mostovoy/Spaldin:2009}. The monopolar magnetic vortex in panel (a) consists of
local spin magnetic moments (black solid arrows) oriented outwards from a point -- note that Maxwell's
equations are not violated; while $\vec{M}$ diverges, it is compensated for
by $\vec{H}$ and so $\vec{B}$ does not diverge. Since the spin moments $\vec{s}_i$  are never
parallel, it is known from the theory of multiferroics that there is a local radial 
electric polarization $\propto \vec{s}_i \times \vec{s}_j$ (unfilled grey arrows)
associated with each pair of spins \cite{Mostovoy:2006,Katsura/Nagaosa/Balatsky:2005}. However these local radial
polarizations are uniform around the vortex and the net electric polarization is zero.
On application of a magnetic field, however, the spin moments reorient to align 
themselves more closely parallel to the field (panel (b)). The local contributions to the
electric polarization no longer average to zero and a net polarization parallel to 
the magnetic field direction results. 

For completeness, we show in Figure~\ref{vortices} (c) and (d) the analogous relationship 
between a toroidal vortex and the off-diagonal magnetoelectric response. In this case an applied
magnetic field modifies the spin orientations so that a net magnetic moment is induced
perpendicular to the direction of applied field.

The remainder of this paper is organized as follows: In the next section we review the
definition of the 
magnetoelectric monopole starting from a multipole expansion of the magnetic field and
show that it couples to the divergence thereof. In Section~\ref{Calculation_and_Measurement}
we describe how the monopole can be calculated from first-principles electronic structure
methods, as well as how it could be directly measured experimentally. We introduce the term 
{\it monopolization} to describe the monopole per unit volume in periodic solids, and show that it 
is natural both theoretically and experimentally to divide the total monopolization into 
two contributions: That arising from the local monopoles around individual ions, and that 
arising from the global distribution of magnetic moments within the solid. We discuss also the 
problems associated with defining the monopolization for an infinite periodic solid, and 
propose a practical solution. In Section~\ref{Transition_metal_phosphates} we present results
of the calculated monopolizations for the family of lithium transition metal phosphates,
LiMPO$_4$, M = Mn, Fe, Co, Ni. All members of this family have the same structure and overall
magnetic order, but they differ in their local magnetic anisotropy and hence their 
magnetic symmetry. We find that the different magnetic symmetries lead to different monopolar
orderings: In one case there is ferromonopolar ordering with a net macroscopic monopolization,
and the remaining three cases have zero net monopolization, but with hidden ``anti-ferromonopolar'' 
orderings that have not previously been identified. 
In section~\ref{GLTheory} we develop the Ginzburg-Landau theory describing the coupling of the 
monopolization to homogeneous external magnetic and electric fields. In the final section we 
discuss the possible relevance of these concepts.

\section{The multipole expansion}
\label{MultipoleExpansion}
Following the derivation in Ref.~\onlinecite{Spaldin/Fiebig/Mostovoy:2008}, we 
consider a magnetization density $\vec{\mu}(\vec{r})$, that may arise from both spin and
orbital contributions, in an inhomogeneous magnetic field $\vec{H}\left(\vec{r}\right)$
that varies slowly on the scale of the system size. Then the interaction energy, 
$H_{\rm int}$, of the magnetization density with the
magnetic field
\begin{equation} \label{eq:inhomog1}
H_{\rm int} = -  \int \vec{\mu}(\vec{r}) \cdot
\vec{H}\left(\vec{r}\right) d^3 \vec{r}
\end{equation}
can be expanded in powers of field gradients calculated at some arbitrary reference point $\vec{r}
= 0$:
\begin{equation} \label{eq:inhomog2}
H_{\rm int} = - \int \vec{\mu}(\vec{r}) \cdot \vec{H}\left(0\right) d^3 \vec{r} 
-  \int r_{i} {\mu}_{j} (\vec{r}) \partial_{i} H_{j}\left(0\right) d^3 \vec{r} - \ldots.
\end{equation}
where $i,j$ are Cartesian directions.
The first term is the interaction of the field with the magnetic moment of the system
\begin{equation}
\vec{m} =  \int \vec{\mu} (\vec{r}) d^3 \vec{r} \quad .
\end{equation}
In the second term, the tensor ${\cal M}_{ij} = \int r_{i} {\mu_j(\vec{r})}d^3 \vec{r}$
with nine components can be decomposed into three parts (summation over repeated indices is
implied):

\begin{itemize}
\item[i)] the pseudoscalar from the trace of the tensor,
\begin{equation} 
a =  \frac{1}{3} {\cal M}_{ii} = \frac{1}{3} \int \vec{r} \! \cdot
\vec{\mu}(\vec{r}) d^3 \vec{r} \quad,
\label{Eqn_monopole} 
\end{equation}
\\

\item[ii)] the toroidal moment vector dual to the antisymmetric part of the tensor, 
$t_{i} = \frac{1}{2} \varepsilon_{ijk} {\cal M}_{jk}$,
\begin{equation} \label{eq:spinT}
\vec{t} = \frac{1}{2}  \int \vec{r} \! \times \vec{\mu}(\vec{r}) d^3 \vec{r} \quad,
\end{equation}
and
\\
\item[iii)] the traceless symmetric tensor $q_{ij}$ describing the quadrupole magnetic moment of
the system,
\begin{eqnarray}
q_{ij} &=& \frac{1}{2}\left({\cal M}_{ij} + {\cal M}_{ji} - \frac{2}{3} \delta_{ij} {\cal
M}_{kk}\right)\nonumber \\ &=& \frac{1}{2} \int \left[r_i \mu_j +
r_j \mu_i - \frac{2}{3} \delta_{ij} \vec{r}\! \cdot \vec{\mu(\vec{r})}
\right] d^3\vec{r} \quad.
\end{eqnarray}
\end{itemize}

The expansion of Eqn. (\ref{eq:inhomog2}) can then be written in the form
\begin{eqnarray} \label{eq:inhomog3}
H_{\rm int} & = & - \vec{m} \cdot \vec{H}\left(0\right) \nonumber \\
            &   & - a \left(\nabla \cdot \vec{H}\right)_{\vec{r} = 0} \nonumber \\
            &   & - \vec{t} \cdot \left[ \nabla \times \vec{H} \right]_{\vec{r} = 0} \nonumber \\
            &   &  - q_{ij} \left(\partial_{i} H_{j} + \partial_{j} H_{i}\right)_{\vec{r} = 0} - \ldots .
\end{eqnarray}
We see that the toroidal moment $\vec{t}$ couples to the curl of the magnetic field,
and the quadrupole moment $q_{ij}$ couples to the field gradient, 
while the pseudoscalar $a$ is coupled to the divergence of magnetic field, and so represents
a monopolar component.

\section{Calculation and measurement of the magnetoelectric monopole in bulk, periodic solids}
\label{Calculation_and_Measurement}

In this section we discuss the difficulties associated with the definition of the monopole
in bulk, periodic solids, and propose solutions that allow a correspondence between calculated
monopole moments and possible experimental measurements. First we note a simplification: Since
the orbital contribution to the magnetization density, $\vec{\mu}^{\text{orb}}(\vec{r})$ is
proportional to $\vec{r} \times \vec{p}(\vec{r})$, where $\vec{p}$ is the momentum, and 
$\vec{r} \cdot \vec{r} \times \vec{p}$ is zero, the orbital contribution to the monopole is
always formally zero, and only the spin contribution need be considered. 

For systems of finite size, such as molecules or molecular clusters, that have zero net magnetic moment, 
the value of the monopole can be evaluated directly from the spin part of the magnetization density through the
integral in Eqn.~\ref{Eqn_monopole}. 
Eqn.~\ref{Eqn_monopole} is not directly applicable to extended
systems where periodic boundary conditions are employed, however, because the integral contains the 
position operator, $\vec{r}$. Therefore for a general continuous magnetization density $\vec{\mu}(\vec{r})$
it will lead to arbitrary values, depending on
the choice of unit cell used in the calculation. 

\subsection{Decomposition of the monopole moment into atomic site and local moment contributions}
\label{Decomposition}

In anticipation of treating the bulk, periodic case, we re-write Eqn.~\ref{Eqn_monopole}
by decomposing the position operator $\vec{r}$ into the positions of the 
constituent atoms, $\vec{r}_\alpha$, relative to some arbitrary origin, plus 
the distance from each atomic center, $[\vec{r} - \vec{r}_\alpha]$. The integral over all space
then separates into a sum over the atomic sites, $\sum_\alpha$ and an integral
around each atomic site, $\int_{\text{as}}$, and Eqn.~\ref{Eqn_monopole} can be rewritten
as
\begin{eqnarray} 
a & = & \frac{1}{3}  \int \vec{r} \! \cdot \vec{\mu}(\vec{r}) d^3 \vec{r} \nonumber \\
  & = & \frac{1}{3}  \sum_\alpha \int_{\text{as}} (\vec{r}_\alpha + [\vec{r}-\vec{r}_\alpha])\! \cdot \vec{\mu}(\vec{r}) d^3 \vec{r} \nonumber \\ 
  & = & \frac{1}{3}  \sum_\alpha \left( \vec{r}_\alpha \cdot \int_{\text{as}} \vec{\mu}(\vec{r}) d^3 \vec{r} + \int_{\text{as}} [\vec{r} - \vec{r}_\alpha] \cdot \vec{\mu}(\vec{r}) d^3 \vec{r} \right) \nonumber \\ 
  & = & \frac{1}{3}  \sum_\alpha \left( \vec{r}_\alpha \cdot \vec{m}_\alpha  + \int_{\text{as}} [\vec{r} - \vec{r}_\alpha] \cdot \vec{\mu}(\vec{r}) d^3 \vec{r} \right) 
\end{eqnarray}
where the summation, $\sum_\alpha$ is over all of the atoms $\alpha$ in the system, and $\vec{m}_\alpha$ is the local
magnetic moment on the $\alpha$th atom.

We see then that the monopole can be decomposed into two components:
The first, comes from the local monopoles at the atomic sites, which arise
from the same current distribution around the site that simultaneously gives rise to the local dipole moment.
We call this contribution $a^{\text{as}}$ for ``atomic site'', and at each site, $\alpha$, it is given by
\begin{equation}
\label{atomic_multipole}
a^\text{as}_{\alpha} = \frac{1}{3} \int_{\text{as}} [\vec{r} - \vec{r}_{\alpha}] . \vec{\mu}(\vec{r}) d^3 \vec{r}
\end{equation}
where the atomic nucleus is at position $\vec{r}_{\alpha}$ and the integral is over some localized region around the atomic nucleus; 
in an electronic-structure calculation this can be chosen to be the ``atomic sphere'' or the ``pseudo-atomic orbital'' depending
on the details of the implementation and the integral can in principle be evaluated over this finite region. 

In practice, we calculate the atomic site contributions to the monopole through expectation values of spherical tensors
using a generalization of the method used previously to obtain inversion-even tensor moments in studies of correlated
$d$ or $f$ electron materials\cite{vanderLaan/Thole:2009,Bultmark_et_al:2009}. 
For each atomic site $\alpha$ a local density matrix $\gamma_{\alpha}$ inside a site-centered sphere is obtained 
from the electronic structure and expanded in spherical harmonics and spinors. In the present work we use the
augmented plane wave plus local orbital (APW+lo) method and these spheres are naturally chosen to be the muffin-tin 
spheres. The density matrices are then further expanded with respect to their behavior (either even or odd) under 
space inversion $i$ and time inversion $\theta$: 
\begin{eqnarray} 
\gamma_{\alpha} & = &\sum_{\nu=0}^{1}\sum_{\eta=0}^{1} \gamma_{\alpha}^{\nu\eta} \nonumber \\
\theta \gamma_{\alpha}^{\nu\eta} & = & (-1)^{\nu}\gamma_{\alpha}^{\nu\eta} \nonumber \\
i \gamma_{\alpha}^{\nu\eta} & = & (-1)^{\eta}\gamma_{\alpha}^{\nu\eta} \quad .
\end{eqnarray}
For magnetoelectrically active multipole moments such as monopoles, only the component that is odd  in both 
space inversion and time reversal that is $\gamma_{\alpha}^{11}$, is relevant. 
In addition, for convenience we expand the density matrices in the Pauli matrices and the identity matrix in spin space,
\begin{align} 
\gamma^{\nu\eta}=&\frac{1}{2}\sum_{\beta=0}^{3} \sigma^{\beta}\gamma_{\alpha}^{\nu\eta\beta} \nonumber \\
\gamma_{\alpha}^{\nu\eta\beta}&=\mathrm{Sp}\, \sigma^{\beta}\gamma^{\nu\eta}\,
\end{align}
where $\mathrm{Sp}$ is the trace over the spin degree of freedom.

\begin{figure}
\centerline{\includegraphics[width=1.10\columnwidth]{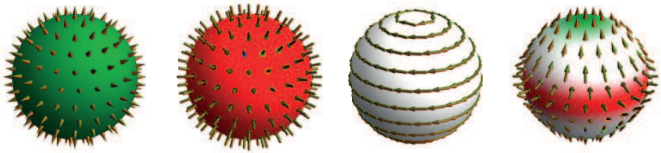}}
\caption{Representation of (left to right) positive and negative monopoles, the $z$ component of the toroidal
moment and the $z^2$ component of the quadrupole moment. } 
\label{hedgehog}
\end{figure}

Now the monopole moment can be written in the form
\begin{align} 
a_{\alpha}=\frac{1}{2} \sum_{\beta={1}}^{3} \mathrm{Tr}\, \Gamma^{(110)} \sigma^{\beta}\gamma_{\alpha}^{11\beta}\,.
\end{align}
Here the operator $\Gamma^{(110)}$ describes the coupling of two rank one tensors, $\bm{r}_{\alpha}$ and $\bm{m}_{\alpha}$, to a rank zero $a_{\alpha}$, and
and $\mathrm{Tr}$ is the trace over the orbital degree of freedom.
In Figure~\ref{hedgehog} we show the generic magnetization textures for positive and negative atomic site
monopoles, as well as for completeness the $z$ component of a toroidal moment and the $z^2$ component of 
the quadrupolar tensor. The arrows represent the magnetization 
orientation on a sphere surrounding an atomic site and the color indicates whether 
the magnetization points outwards (green) or inwards (red).

Note that these atomic site monopoles can in principle be measured by resonant 
x-ray spectroscopy\cite{Lovesey/Scagnoli:2009}, which has been used successfully to 
detect an atomic site toroidal moment\cite{Staub_et_al:2009,Scagnoli_et_al:2011}. 
No unambiguous measurement of atomic monopoles has been made to date, however, because 
a material has not yet been identified that meets the stringent conditions required to achieve an observation
in the resonant x-ray measurement.  We point out also that, provided that the local 
magnetic site is not an inversion center, the atomic monopoles can be non-zero 
even in a system with overall zero monopole moment; we will explore some examples in
Section~\ref{Transition_metal_phosphates}. Such systems might be described as 
``anti-monopolar'' and should show a $q$-dependent magnetoelectric effect. 

The second contribution to the monopole, which we write $a^{\text{lm}}$ for ``local moment'', arises from 
representing the magnetization density by a distribution of localized magnetic moments $\vec{m}_\alpha$ at 
the atomic sites:
\begin{equation}
\label{a_local_moment}
a^\text{lm} = \frac{1}{3} \sum_\alpha \vec{r}_\alpha . \vec{m}_\alpha
\quad .
\end{equation}
In systems such as insulating $3d$ transition metal oxides, which have large localized magnetic
moments that are spatially separated by distances of a few \AA\, we expect this contribution to
be the dominant contribution to the total monopole.

\begin{figure}
\centerline{\includegraphics[width=0.8\columnwidth]{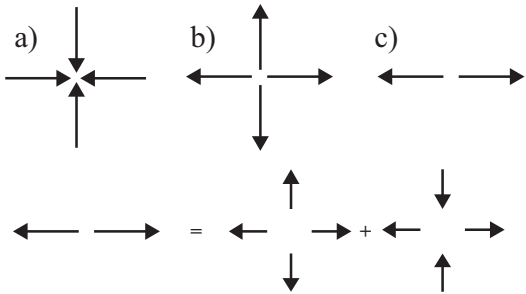}}
\caption{Representative arrangements of local magnetic moments (shown by arrows) that
have monopolar contributions. The arrangements in a) and b) are purely monopolar, and
have equal and opposite monopoles. c) consists of the sum of a monopolar contribution (of size
half that of b) and a quadrupolar contribution; the decomposition is shown in the lower
panel.}
\label{monopoles_cartoons}
\end{figure}

Using Eq.~(\ref{a_local_moment}) we can straightforwardly evaluate the
monopoles of the arrangements of magnetic moments shown in
Fig.~\ref{monopoles_cartoons}.  Taking the $\pm y$-oriented magnetic moments
to be spaced a distance $d$ apart along the $y$ direction, and the
$\pm x$-oriented moments a distance $d$ apart along $x$, 
then the monopoles of arrangements a) and b) in Fig.~\ref{monopoles_cartoons} are 
$a = -\frac{2}{3}d m$ and $+\frac{2}{3}d m$ respectively, 
where $m$ is the magnitude of each local magnetic dipole moment. 
Applying Eq.~\ref{a_local_moment} to the arrangement show in c) yields the value 
$+\frac{1}{3}d m$; this can also be obtained by inspection by recognizing that
c) consists of a monopole with magnetic moments at the same position of as in b)
but of half the magnitude, plus a quadrupole, as shown in the lower panel of 
Fig.~\ref{monopoles_cartoons}.

The total monopole resulting from these two contributions is then
\begin{equation}
a =  a^\text{lm}  + \sum_\alpha a_\alpha^\text{as} 
\end{equation}
where the sum is over all the atomic sites.

In all the cases shown in Fig.~\ref{monopoles_cartoons}, the net magnetization is zero. 
There exists a complication, however, in the case where the region over which the monpole 
is to be evaluated has a net magnetic dipole. The complication is that
all multipoles in systems with non-zero lower-order multipoles (the magnetic dipole in the case
of the magnetoelectric monopole) are dependent on the choice of origin used to evaluate them. 
It is straightforward to see that for systems with nonvanishing magnetic dipole moment, 
for a change of origin defined by 
\begin{equation}
\vec{r} \rightarrow \vec{r}' = \vec{r} + \vec{R}_0
\end{equation} 
the monopole changes as 
\begin{equation}
a \rightarrow a' = a + \frac{1}{3} \vec{R}_0 . \int \vec{\mu(r)} d^3 \vec{r} \quad .
\end{equation}
It remains an open question in general, which we do not address here, whether such origin dependence of 
the multipoles is physically
meaningful (see for example Ref.~\onlinecite{Visschere:2006}).  One practical approach is to always 
choose as the origin the position of the average magnetic moment, 
$\bar{\vec{R}}$, defined so that 
$ \int \vec{\mu(r-\bar{R})} d^3 \vec{r} =0 .  $
This is equivalent to neglecting any
uncompensated part of the magnetization and retaining only the compensated part in the calculation
of the monopole\cite{Ederer/Spaldin:2007}. Care must be taken, however, in situations where a
change in net magnetic dipole moment, or a structural rearrangement occurs, to ensure that a 
consistent choice of origin is maintained.

\subsection{Bulk systems with periodic boundary conditions; the problem of multi-valuedness}
\label{sec:bulk}

Next we turn to the case of a system with periodic boundary conditions.
It is often convenient to describe the properties of a bulk crystalline solid in terms
of a small repeat unit -- the unit cell -- which is then replicated using periodic
boundary conditions to generate the infinite solid. Many intensive quantities such as
the magnetization, which is defined to be the magnetic moment per unit volume,
can then be simply obtained as the value of the quantity in a single unit cell divided by the
unit cell volume. 
For the case of the macroscopic monopole per unit volume -- which we propose to call the 
{\it monopolization} by analogy with magnetization, polarization, etc. -- Eqn.~\ref{Eqn_monopole} 
is not directly applicable to extended systems with periodic boundary
conditions, because for a general continuous magnetization
density $\vec{\mu}(\vec{r})$, Eq.~(\ref{Eqn_monopole}) evaluated over one unit
cell will lead to arbitrary values, depending on the particular choice
of unit cell used in the calculation. We note that this behavior is distinct from the 
origin dependence discussed in Section~\ref{Decomposition}, and persists even in
the case when the net magnetization is zero. In fact
the difficulties are exactly analogous to those encountered in defining a macroscopic
bulk toroidization, and indeed reflect those involved in defining a macroscopic bulk ferroelectric 
polarization, which were solved through the introduction of the modern theory
of polarization\cite{Resta:1993,Resta:1994,King-Smith/Vanderbilt:1993}. 
A proposed solution in the case of the toroidization was described
in detail in Ref.~\onlinecite{Ederer/Spaldin:2007}. In this section we extend the description 
to the case of the monopole and address the following questions:
\begin{enumerate}
\item{How should the monopole density -- the monopolization -- of a bulk periodic solid be formally defined?}
\item{What are the consequences of the periodic boundary conditions within a bulk crystalline solid?}
\end{enumerate}

For simplicity we develop the formalism for the case of the monopolization coming from the 
local moment contribution. First we note that, as we shall see later, the formalism requires 
that each local 
moment, $\vec{m}_\alpha$, is equal to an integer number of Bohr magnetons. Since we consider
only the spin part of the magnetic moment (the orbital part does not contribute to the monopole),  
a magnetic moment that is an integer number of Bohr magnetons corresponds to the moment of an
integer number of electrons. 
In general, however, an integer number is not obtained from integrating the magnetization density 
over a sphere around an atomic site in a solid; in fact this number is not uniquely defined as it depends
on the choice of integration radius. Rather, the spin moment of the corresponding spin-polarized Wannier 
function should be used; since a Wannier function in an insulating system contains an integer number of 
electrons its spin is always an integer number of Bohr magnetons.

We then define the local moment monopolization,
$A^\text{lm} =
a^\text{lm}/V$, where $V$ is the volume of the system with local moment monopole 
$a^\text{lm}$.  Then, for a large finite system containing $N$ identical
unit cells each of volume $\Omega$:
\begin{align}
  A^\text{lm} & = \frac{1}{3 N \Omega} \sum_{\alpha}
  \vec{r}_{\alpha} . \vec{m}_{\alpha} \\ & = \frac{1}{3 N \Omega}
  \sum_{n,i} (\vec{r}_i+ \vec{R}_n) . \vec{m}_i \quad .
\end{align}
Here, $\vec{r}_i$ are the positions of the magnetic moments
$\vec{m}_i$ relative to the same (arbitrary) point within each unit
cell, $\vec{R}_n$ is a lattice vector with index $n$, and we have
used the fact that the orientation of the magnetic moments is the same
in each unit cell. The summation over $i$ indicates the summation over
all moments within a unit cell, and that over $n$ indicates the
summation over all unit cells.  Expanding the scalar product, we
obtain:
\begin{align}
A^\text{lm} & = \frac{1}{3 \Omega} \sum_i \vec{r}_i . \vec{m}_i +
\frac{1}{3 N \Omega} \sum_{n} \vec{R}_n . \sum_i \vec{m}_i \nonumber \\ 
& = \frac{1}{3 \Omega} \sum_i \vec{r}_i . \vec{m}_i \quad ,
\end{align}
using the fact that the sum over all lattice
vectors contains both $\vec{R}_n$ and $-\vec{R}_n$, so that $\sum_n \vec{R}_n = 0$. Thus,
the local moment monopole of a system of $N$ unit cells is just $N$ times
the monopole evaluated for one unit cell, and the corresponding
monopolizations are identical.

In an infinite periodic solid, we have a freedom in choosing the basis
corresponding to the primitive unit cell of the crystal. In
particular, we can translate any spin of the basis by a lattice vector
$\vec{R}_n$ without changing the overall periodic
arrangement. However, such a translation of a spin by
$\vec{R}_n$ leads to a change in the local moment monopolization as
follows:
\begin{equation}
\label{quantum}
\Delta A^{\text{lm}}_{ni} = \frac{1}{3 \Omega} \vec{R}_n \cdot \hat{m}_i \mu_B \quad ,
\end{equation}
where $\hat{m}_i$ is a unit vector oriented in the direction of magnetic
moment $\vec{m}_i$.
The freedom in choosing the basis corresponding to the primitive unit
cell thus leads to a multivaluedness of the monopolization with respect
to certain ``increments'' (defined by Eq.~({\ref{quantum})) for each
magnetic sub-lattice $i$ and lattice vector $\vec{R}_n$.

This multivaluedness of the monopolization is reminiscent of the modern theory of electric
polarization,\cite{King-Smith/Vanderbilt:1993,Vanderbilt/King-Smith:1993,Resta:1994}
where the polarization changes by $e\vec{R}_n/\Omega$ when
an elementary charge $e$ is translated by a lattice vector
$\vec{R}_n$. The resulting multivaluedness has led to the concept of
the ``polarization lattice'' corresponding to a bulk periodic
solid,\cite{Vanderbilt/King-Smith:1993} with $e\vec{R}_n/\Omega$
called the ``polarization quantum'' if $\vec{R}_n$ is one of the three
primitive lattice vectors. An even closer analogy is provided by the toroidization,
which is multivalued with values spaced by the toroidization increment 
$\frac{1}{2\Omega} \vec{R}_n \times \vec{m}$, corresponding to translation of an
elementary magnetic moment, $\vec{m}$ by a lattice vector \cite{Ederer/Spaldin:2007}.
Eq.~(\ref{quantum}) suggests the existence of an analogous ``monopolization lattice'', 
with monopolization increments $\frac{1}{3 \Omega} \mu_B \vec{R}_n \cdot \hat{m}_i$, where $\vec{R}_n$ is any
primitive lattice vector and $\hat{m}_i$ are the unit vectors indicating the orientations of 
the magnetic moments.
Note that the monopolization, and hence the monopolization increments are scalar
quantities. As a result the corresponding monopolization lattice can become rather
dense, particularly in cases where the three lattice vectors are unequal but close in
size, and the spin moments are noncollinear and canted away from the lattice vector directions. 

We illustrate the behavior and implications of the monopolization lattice next
with a simple model one-dimensional example. 

\subsection{A one-dimensional example}
\label{chain}

\paragraph{The periodic non-monopolar state.}
\label{afm_chain_a}

\begin{figure}
\centerline{\includegraphics[width=1.0\columnwidth]{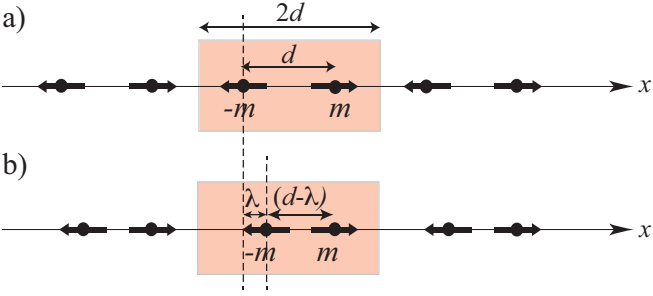}}
\caption{Calculation of the monopolization for two different
  one-dimensional antiferromagnetic periodic arrangements of magnetic
  moments. Our choice of unit cell is indicated by the shaded area in
  each case. a) shows a non-monopolar state, which is space-inversion
  symmetric with respect to each moment site. b) is a monopolar 
  state.}
\label{lattice}
\end{figure}

To illustrate some consequences of the multivaluedness of the
monopolization in periodic systems described in the previous section,
we now consider the example of a one-dimensional antiferromagnetic
chain of equally spaced magnetic moments as shown in
Fig.~\ref{lattice}a. The moments, with magnitude $m = \mu_B $, are spaced a
distance $d$ apart from each other along the $x$ axis, and are
alternating in orientation along $\pm x$. Thus, the unit cell length
is $2d$ and there are two oppositely oriented magnetic moments in each
unit cell. Since this configuration does not possess a macroscopic
magnetic dipole moment, the corresponding monopole moment is
origin independent. 

The arrangement of magnetic moments in Fig.~\ref{lattice}a is
space-inversion symmetric with respect to each moment site and thus
cannot exhibit a macroscopic monopole moment. 
The local moment monopole of the single unit cell highlighted in
Fig.~\ref{lattice}a, calculated using Eq.~(\ref{a_local_moment}), however,
is identical to that calculated for the finite moment configuration in
Fig.~\ref{monopoles_cartoons}c, i.e. $a^\text{lm} = \frac{1}{3} d m$, and 
the corresponding monopolization, $A^\text{lm} =
a^{\text{lm}}/\Omega = \frac{1}{3} \frac{d m}{2d} = \frac{1}{3} \frac{m}{2}$ 
(since the ``volume''
$\Omega$ of the one-dimensional unit cell is just its length, $2d$).
Since the moments of magnitude $\mu_B$ are oriented exactly parallel to the
$x$ axis, the elementary monopolization increment in this case is $\Delta A^\text{lm}
= \pm \frac{1}{3} \mu_B $, which means that the monopolization 
of the unit cell is exactly equal to one half of the monopolization 
increment, and the allowed monopolization values for the periodic
arrangement are $A_n = (\frac{1}{2} + n)\frac{1}{3} \mu_B$, 
where $n$ can be any integer number.

We see that in our example the allowed local moment monopolization values form a
one-dimensional lattice of values, centrosymmetric around the
origin. This is analogous to the cases of the electric polarization and
the toroidization,
where the polarization and toroidization lattices are invariant under all symmetry
transformations of the underlying crystal structure. In particular,
the polarization and toroidization lattices corresponding to centrosymmetric crystal
structures are inversion symmetric, which is achieved in lattices that include
either the zero or the half quantum/increment. 
We see that the same holds true for the local moment monopolization
of our one-dimensional example, and that a centrosymmetric set of
monopolization values can be understood as representing a non-monopolar
state of the corresponding system. We also note that the formalism is only
consistent for the case of local magnetic moments corresponding to integer numbers 
of Bohr magnetons, which in turn correspond to the spin contribution from integer
numbers of electrons.

In the case of the electric polarization, it is now widely recognized that only 
differences in the polarization lattices between different configurations, such
as between a centrosymmetric non-polar reference structure and 
a ferroelectric polar crystal, are in fact measurable quantities.
Since these differences are the same for each point of the
polarization lattice they are well-defined quantities.  
Likewise in the case of the toroidization, only differences in toroidization lattices
between for example different arrangements of magnetic moments or different ionic
positions are measurable\cite{Ederer/Spaldin:2007}.
In the next section we show that, in analogy with the cases of the toroidization
and electric polarization,
only differences in local moment monopolization, corresponding to two different bulk
configurations, are measurable quantities and correspond to physical observables
such as the difference in monopolization between a ferromonopolar state and its
non-monopolar paraphase. Such quantities
can be obtained by monitoring the change in monopolization on one
arbitrarily chosen \emph{branch} within the allowed set of values,
when transforming the system from the initial to the final state along
a well-defined path.

\paragraph{Monopolar state and changes in monopolization.}

In order to obtain a nontrivial macroscopic monopolization the
system has to break both space and time inversion symmetry. In the
case of the one-dimensional antiferromagnetic chain this can be
achieved by ``moment pairing'', i.e.  if the distances between
neighboring magnetic moments alternate as shown in
Fig.~\ref{lattice}b.  Here the magnetic moments of magnitude $m= \mu_B$ are
spaced alternately a distance of $(1-\lambda)d$ and $(1+\lambda)d$ apart from each
other along the $x$ axis ($-1 < \lambda < 1$). 
The non-monopolar example above corresponds
to $\lambda = 0$.  Since the unit cell size is the same as in the
non-monopolar case, the elementary monopolization increment is again
$\Delta A^\text{lm} = \pm \frac{1}{3} \mu_B$. The monopolization of
the unit cell indicated in Fig.~\ref{lattice}b is 
$A^\text{lm} = \frac{1}{3} (-\frac{\lambda}{d} +1) \frac{\mu_B}{2}$, so that the 
allowed values of
$A^\text{lm}$ for the full periodic arrangement are:
\begin{equation}
\label{T-chain}
A^\text{lm} = \left(\frac{1}{2} +  n \right) \frac{1}{3} (-\frac{\lambda}{d} +1) \mu_B
\quad .
\end{equation}
Fig.~\ref{a_of_d} shows the allowed monopolization values as a function
of the displacement $\lambda$ of the moments from their positions in the
centrosymmetric, non-monopolar state.

\begin{figure}
\centerline{\includegraphics[width=0.9\columnwidth]{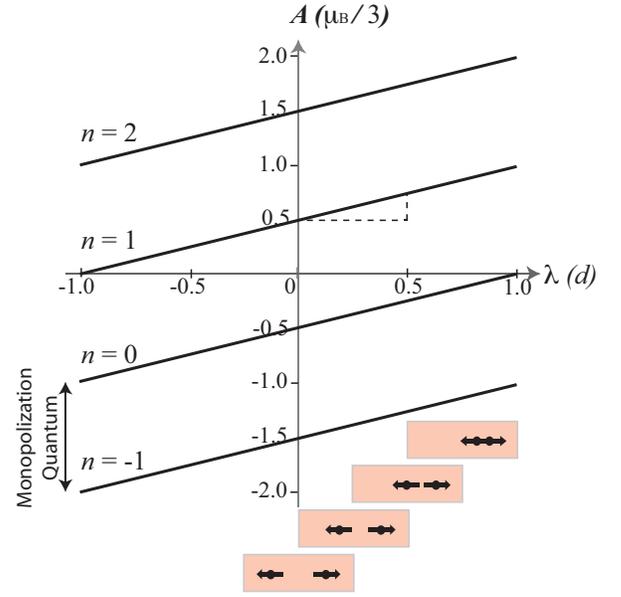}}
\caption{Allowed values of the monopolization for the antiferromagnetic
  chain of Fig.~\ref{lattice} as a function of displacement $\lambda$ from
  the non-toroidal case ($\lambda=0$). The cartoons at the bottom indicate
  the corresponding positions of the magnetic moments within the unit
  cell. }
\label{a_of_d}
\end{figure}

The change in monopolization between two configurations with $\lambda=\lambda_1$
and $\lambda=\lambda_2$ for a certain branch $n$ is given by:
\begin{equation}
A_n^{\text{lm}}(\lambda_2) - A_n^{\text{lm}}(\lambda_1) = \frac{1}{3} 
\frac{\lambda_2 - \lambda_1}{d} \frac{\mu_B}{2}
\quad ,
\label{A_model}
\end{equation}
i.e. it is independent of the branch index $n$. In particular, if the
non-centrosymmetric distortion is inverted ($\lambda_2=\lambda_0$, $\lambda_1=-\lambda_0)$, the
change in monopolization is $2A^\text{lm}_\text{s} = \frac{1}{3} \frac{\lambda_0 \mu_B}{d}$ so that
$A^\text{lm}_\text{s} = \frac{1}{3} \frac{\lambda_0 \mu_B}{2d}$ 
can be interpreted as the
\emph{spontaneous monopolization}, again in analogy to the case of the
electric polarization, where the spontaneous polarization is given by
the branch-independent change in polarization compared to a
centrosymmetric reference structure.

Another possible way to alter the monopolization is by changing the
orientation of the magnetic moments instead of changing their
positions. In particular, we expect that a full 180$^\circ$ rotation
of all magnetic moments, which is equivalent to the operation of time
reversal, should invert the macroscopic ``spontaneous monopolization'',
and should therefore lead to the same change $2A_\text{s}^{\text{lm}}$ as
discussed above. If we allow the magnetic moments to rotate out of the
$x$ direction, while preserving the antiparallel alignment of the two
basis moments, the monopolization is given by
\begin{equation}
A^\text{lm}_n(\lambda,\alpha) = \left(\frac{1}{2} +  n \right) 
\frac{1}{3} (-\frac{\lambda}{d} +1) \mu_B \cos{\alpha}
\quad ,
\end{equation}
where $\alpha$ is the angle between the magnetic moments and the $x$
direction. Note here a difference from the case of the toroidization -- 
since the monopolization is a scalar, rotation of the magnetic moments
away from perfect alignment reduces the absolute magnitude of the
monopolization. In contrast, in the toroidal case a rotation could reduce the
toroidization along one axis while simultaneously increasing it along
another.
Interestingly, in this example, the magnetic moment rotation which reduces the
monopolization induces a toroidization, effectively converting the monopolar
response into a toroidal one through the moment reorientation. 
The change in monopolization for a full 180$^\circ$ rotation
of the moments is thus:
\begin{equation}
\label{A-change}
A^\text{lm}_n(\lambda_0,180^\circ) - A^\text{lm}_n(\lambda_0,0^\circ) 
 = 2 \left(\frac{1}{2} +  n \right) 
\frac{1}{3} (-\frac{\lambda_0}{d} +1) \mu_B
\quad ,
\end{equation}
and apparently depends on the branch index $n$. However, if one
calculates the same change in monopolization for the non-monopolar state
with $d=0$, one obtains:
\begin{equation}
\label{improper}
A^\text{lm}_n(0,180^\circ) - A^\text{lm}_n(0^\circ) 
 = -2 \left(\frac{1}{2} +  n \right) 
\frac{1}{3} \mu_B
\quad .
\end{equation}
Obviously, in this case the corresponding change in macroscopic
monopolization should be zero, since both the initial and final states
(and all intermediate states) correspond to a non-monopolar
configuration and thus $A_\text{s}^{\text{lm}}=0$. If one subtracts the \emph{improper}
change in $A^{\text{lm}}$, Eq.~(\ref{improper}), from the change in
monopolization calculated in Eq.~(\ref{A-change}), one obtains the
\emph{proper} change in monopolization 
$2A^\text{lm}_\text{s} = \frac{1}{3} \frac{\lambda_0 \mu_B}{d}$,
which is identical to that obtained by inverting the non-centrosymmetric
distortion $\lambda$. Here, we use the terminology ``proper'' and
``improper'' in analogy to the case of the proper and improper
piezoelectric response, \cite{Vanderbilt:2000} where a similar branch
dependence is caused by volume changes of the unit cell, and the
improper piezoelectric response has to be subtracted appropriately.

\begin{figure}
\centerline{\includegraphics[width=0.95\columnwidth]{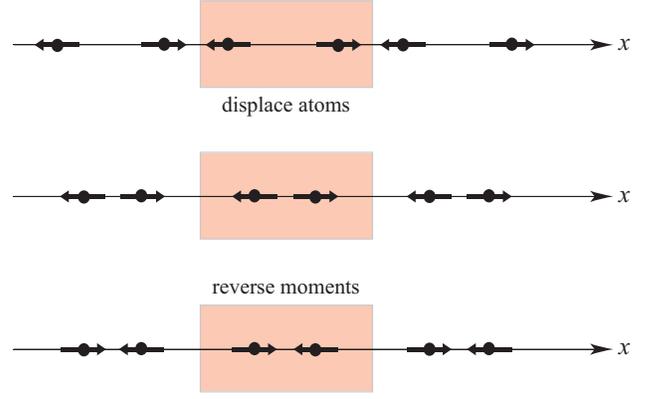}}
\caption{Effect on the magnetic moment configuration of
Fig.~\ref{lattice}b (middle panel) of a reversal of all magnetic moments
(lower panel) and of a reversal of the non-centrosymmetric distortion $d$
(upper panel). Note that the upper and lower final states are identical, with
the moments in the upper and lower panels translated by half a unit cell 
relative to each other.}
\label{spin_flip}
\end{figure}

Fig.~\ref{spin_flip} shows the initial and final states for the two
cases where either the atoms carrying the moments are displaced, or the magnetic moment
directions are inverted. The two final states are equivalent except
for a translation of all moments by half a unit cell along $y$, which,
due to Neumann's principle, is irrelevant for the macroscopic
properties. The spontaneous monopolization of the upper state in
Fig.~\ref{spin_flip} is therefore the same as for the lower state in the
Figure.

\section{Monopolizations in real materials -- the {Li} transition-metal phosphates}
\label{Transition_metal_phosphates}

\begin{figure}
\centering
\def\svgwidth{\columnwidth}
\centerline{\includegraphics[width=0.95\columnwidth]{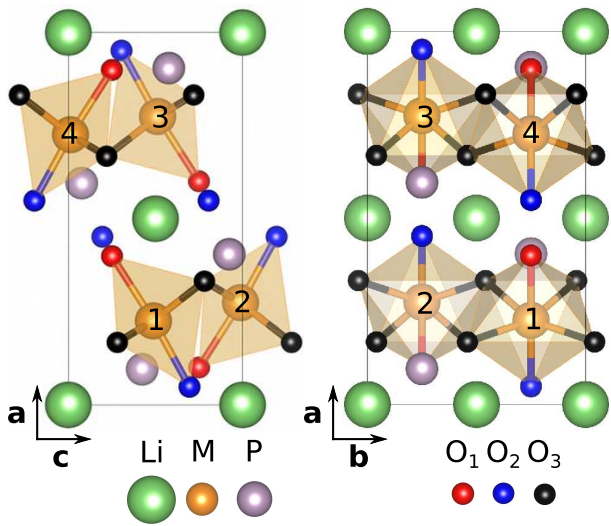}}
\caption{\label{fig6} Structure of the lithium transition metal phosphates. The 1 - 4 
labeling of the transition metal atoms is consistent with their labeling in 
Tables~\ref{tab:magnetic_orderings} and~\ref{tab:LiMPO4-character}.
}
\end{figure}

We now turn to a real materials example, and choose the family of lithium
transition-metal phosphates, LiMPO$_4$, M = Mn, Fe, Co, Ni, as our model
system. All of the LiMPO$_4$ compounds crystallize in the olivine structure with the orthorhombic 
space group $Pnma$ and the crystallographic point group 
$D_{2h}$.\cite{Destenay:1950,Newnham/Redman:1965,Geller/Durand:1960,Santoro/Newnham:1967,Abrahams/Easson:1993} 
The lattice parameters and atomic coordinates, obtained from first-principles calculations in
this work and Refs.~\onlinecite{Bousquet/Spaldin/Delaney:2011} and \onlinecite{Scaramucci_et_al:2012}, 
are given in Table~\ref{Structure}.

\begin{table}
\centering
\begin{tabular}{l l l S S S S}
\hline
\hline
       &     &     & Mn    & {Fe}  & Co  & {Ni}     \\ \hline
a (\AA)&     &     &10.440 &10.330 & 10.202& 10.032 \\
b/a    &     &     & 0.583 &0.582 & 0.581  &0.584   \\
c/a    &     &     & 0.455 & 0.454 &0.461 & 0.466   \\ \hline
M      & $4c$&x    & 0.280 & 0.282& 0.223 &  0.225  \\
M     & $4c$ &z    & 0.477 & 0.480&0.507 &  0.488   \\
P     & $4c$ &x    & 0.093 & 0.096& 0.096&  0.095   \\
P     & $4c$ &z    &-0.085 & -0.072&-0.074& -0.076  \\
O$_1$ & $4c$ &x    & 0.097 & 0.097&0.101 &  0.101   \\
O$_1$ & $4c$ &z    & 0.237 & 0.254&0.248 &  0.250   \\
O$_2$ & $4c$ &x    & 0.455 & 0.458&0.455 & 0.452    \\
O$_2$ & $4c$ &z    &-0.292 & -0.300&-0.193 &-0.305  \\
O$_3$ & $8d$ &x    & 0.171 & 0.168&0.168 &  0.170   \\
O$_3$ & $8d$ &y    & 0.048 & 0.045& 0.457 &  0.040  \\
O$_3$ & $8d$ &z    &-0.218 & -0.204&-0.212 & -0.220  \\ \hline
\end{tabular}
\caption{a, b and c lattice parameters and Wyckoff positions for the lithium transition
metal phosphates, Li$M$PO$_4$, $M$ = Mn, Fe, Co and Ni. All values were obtained by 
structural relaxation using density functional theory within the LSDA$+U$ method as
described in the text.}
\label{Structure}
\end{table}

The transition metal cations occupy the sites with Wyckoff positions $4c$; these are surrounded by
strongly distorted oxygen octahedra and have local $C_s=\{e,i_{2y}\}$ symmetry. All compounds have 
a transition to an 
antiferromagnetic state at some tens of kelvin. The resulting magnetic order breaks the inversion symmetry 
in all cases and hence allows for the linear magnetoelectric effect. Across the series, however, three distinct 
antiferromagnetic orderings emerge
\cite{Vaknin_et_al:2002,Liang_et_al:2008,Santoro/Newnham:1967,ToftPetersen_et_al:2012,Jensen_et_al:2009}, 
summarized in Table~\ref{tab:magnetic_orderings}. These different antiferromagnetic
orderings lead in turn to different magnetic symmetries and different allowed monopolar contributions. 

\begin{table}
\centering
\begin{tabularx}{\columnwidth}{X c c  c}
\hline
\hline
        & Mn   & Fe / Co    & Ni   \rule[-1ex]{0pt}{3.5ex} \\
\hline
$m_1$   & ($m$, 0, 0)  & (0, $m$, 0)   & (0, 0, $m $)    \rule[-1ex]{0pt}{3.5ex}\\
$m_2$   & ($-m$, 0, 0) & (0, $-m$, 0)  & (0, 0, $-m$)   \rule[-1ex]{0pt}{3.5ex}\\
$m_3$   & ($-m$, 0, 0) & (0, $-m$, 0)  & (0, 0, $-m$)    \rule[-1ex]{0pt}{3.5ex}\\
$m_4$   & ($m$, 0, 0)  & (0, $m$, 0)   & (0, 0, $m $)     \rule[-1ex]{0pt}{3.5ex}\\ \hline
$|m|^{\text{lm}}_\text{spin}$ ($\mu_B$) & 5  & 4 / 3  & 2 \\ \hline
\end{tabularx}
\caption{Experimentally determined magnetic orderings for the lithium transition metal phosphates.
For simplicity we neglect small cantings of the magnetic moments away from the easy axis that are 
reported or known for many of the compounds. We also list the local-moment spin magnetic moment
for each transition metal ion.
}
\label{tab:magnetic_orderings}
\end{table}

\begin{table*}
\centering
\begin{tabular}{|l|rrrrrrrr|c|c|c|c|c|c|c||c|c|}
\hline
\multicolumn{9}{|c|}{point group}&\multicolumn{7}{c||}{$4c$ (M)}&$4a$ (Li) &$8d$ (O$_3$)\\\hline
$D_{2h}$ & $e$ & $c_{2z}$ & $c_{2y}$ &$c_{2x}$& $i$ & $i_{2z}$ & $i_{2y}$ &$i_{2x}$ & $a$, $q_{z^{2}/x^{2}-y^{2}}$ & $t_{x}$, $q_{yz}$& $t_{y}$, $q_{zx}$& $t_{z}$, $q_{xy}$& $m_{x}$& $m_{y}$& $m_{z}$&$a$&$a$\\
\hline
$A_{g}$ & 1 & 1 & 1 & 1 & 1 & 1  & 1 & 1 & 0 & $+--+$ & 0& $++--$& 0 & $+-+-$ & 0&0&$++++----$\\
$B_{1g}$ & 1 & 1&  -1 & -1 & 1 & 1 & -1& -1   &$++--$ & 0 &$+--+$ & 0& $+-+-$ & 0 & $++++$&0&$++----++$\\
$B_{2g}$ & 1 & -1&  1 & -1 & 1 & -1 & 1& -1  & 0  &  $++--$& 0&$+--+$ & 0 & $++++$ & 0&0&$+-+--+-+$\\
$B_{3g}$ & 1 & -1&  -1 & 1 & 1 & -1 & -1& 1   & $+--+$& 0 &$++--$ & 0 & $++++$ & 0 &$+-+-$&0&$+--+-++-$\\
$A_{u}$ & 1 & 1 & 1 & 1 & -1 & -1  & -1 & -1  & $++++$& 0 &$+-+-$ & 0 & $+--+$ & 0& $++--$&$++++$&$++++++++$\\
$B_{1u}$ & 1 & 1&  -1 & -1 & -1 & -1 & 1& 1  & 0& $+-+-$ & 0& $++++$&0&$+--+$&0&$++--$&$++--++--$\\
$B_{2u}$ & 1 & -1&  1 & -1 & -1 & 1 & -1& 1   & $+-+-$& 0 &$++++$ & 0 &$++--$ &0&$+--+$&$+-+-$&$+-+-+-+-$\\
$B_{3u}$ & 1 & -1&  -1 & 1 & -1 & 1 & 1& -1  &  0 & $++++$ & 0& $+-+-$ & 0 & $++--$& 0&$+--+$&$+--++--+$\\
\hline
\end{tabular}
\caption{Character table of the $D_{2h}$ point group, and symmetry analyses for the $4c$ site (dipole, monopole, toroidal and quadrupole
ordering) and the $4a$ and $8d$ sites (monopole ordering only) of the Pnma space group.}
\label{tab:LiMPO4-character}
\end{table*}

\subsection{Symmetry analysis}
In Table~\ref{tab:LiMPO4-character} we show the character table of the $D_{2h}$ symmetry group and indicate which irreducible 
representations are adopted by each possible collinear ordering of the transition metal magnetic
moments, $m$, along the cartesian axes, as well as the symmetries 
of the possible monopolar $a$, toroidal $t$ and quadrupolar $q$ orderings on the transition metal sites. 

In LiMnPO$_4$ the easy axis is the $a$ axis, and the magnetic moments adopt a C-type antiferromagnetic ordering with order 
parameter $m_1 - m_2 - m_3 + m_4$\cite{ToftPetersen_et_al:2012},
this combination belongs to the $A_u$ irreducible representation of the the $D_{2h}$ symmetry group. (This ordering allows for a simultaneous 
A-type antiferromagnetic canting along the $c$ axis which is negligible in our DFT calculations and we neglect here.
Note that a weak ferromagentic canting has also been reported, which is not compatible with the $Pnma$ symmetry
analysis\cite{Arcon_et_al:2004}; this we also neglect.) 
We see from the line corresponding to the $A_u$ irreducible representation in Table~\ref{tab:LiMPO4-character} that the ordering of local M-site monopole moments all with the same sign also has 
$A_u$ symmetry, therefore LiMnPO$_4$ is ferromonopolar and supports a macroscopic monopolization. Conversely there is no net 
toroidal moment, with only an anti-ferrotorodial ordering along the $b$ direction allowed on the Mn sites. This is consistent with the experimental 
observation that the magnetoelectric response has only diagonal components\cite{Mercier/Gareyte/Bertaut:1967}. We note also that the $z^2$ and $x^2-y^2$ quadrupolar 
components have the same symmetry as the monopole; these quadrupolar contributions are responsible for the inequality between the 
magnitudes of the diagonal elements of the magnetoelectric tensor. 

LiCoPO$_4$ has been of particular recent interest because the observation of ferrotoroidic domains using nonlinear optical techniques 
has been reported.\cite{VanAken_et_al:2007} Both LiCoPO$_4$ and LiFePO$_4$ also adopt a C-type antiferromagnetic ordering, but in contrast 
to LiMnPO$_4$, both have their easy axis primarily along the $b$ axis\cite{Santoro/Segal/Newnham:1966,Liang_et_al:2008}. 
This corresponds to the $B_{1u}$ irreducible representation which we see from
Table~\ref{tab:LiMPO4-character} disallows both a macroscopic monopolization and any local monopolar contribution on the transition metal 
sites. This symmetry allows, however, a toroidal moment parallel to the $c$ axis. As a result the magnetoelectric responses of both compounds are 
entirely off-diagonal\cite{Mercier/Bauer/Fouilleux:1968,Mercier/Gareyte/Bertaut:1967}}, although $\alpha_{xy}$ is not exactly equal to -$\alpha_{yx}$ (which would be the case for a purely toroidal response) 
because a ferroquadrupolar $q_{xy}$ component is allowed with the same symmetry as $t_z$. 
(We note that recently it was found that the magnetic moments in LiCoPO$_4$ and LiFePO$_4$ are 
rotated slightly away from the $b$ direction \cite{Vaknin_et_al:2002,Li_et_al:2006}. Such a symmetry lowering
is not compatible with the $Pnma$ space group and requires an additional structural distortion that has not
yet been identified. We do not treat these further symmetry lowerings here.)

Finally we turn to the case of LiNiPO$_4$, which again has C-type AFM ordering, but this time with easy axis along the $c$ direction\cite{Jensen_et_al:2009}, 
so that the Ni sublattice has magnetic point group $mm'm$ and transforms according to the $B_{2u}$ representation. (This symmetry also
allows a small A-type AFM canting of the magnetic moments along the $a$ direction which has been reported \cite{Jensen_et_al:2009} and which we 
neglect here). While this symmetry 
does not allow a net macroscopic monopolization, local monopoles are allowed on the Ni ions and must order with an {\it antimonopolar} arrangement. 
A macroscopic toroidal moment is again allowed, this time along the $b$ direction, consistent with the 
corresponding off-diagonal magnetoelectric effect\cite{Mercier/Bauer:1968,Jensen_et_al:2009,Bousquet/Spaldin/Delaney:2011}. 

In this series, therefore, we find one example -- LiMnPO$_4$ -- of a material with a net monopolization, in which the local monopole 
moments on the transition metal sites are aligned in a ferromonopolar arrangement. We also find an example -- LiNiPO$_4$ -- which has 
no macroscopic monopolization, but has a finite-$q$ antimonopolar ordering on the transition metal sites. In the remaining two compounds 
-- LiCoPO$_4$ and LiFePO$_4$ -- the macroscopic monopolization and the local monopoles on the transition metal sites are both zero by 
symmetry. We summarize our symmetry analysis in Table~\ref{Eric_Symmetry}.

\begin{table}
 \begin{center}
 \begin{small}
\begin{tabular}{lccccc}
 \hline
\hline
label    & $M$    & magnetic order &  ME & Toroidal & Monopole \rule{0pt}{0.4cm}\\
\hline
$A_u$	 & Mn     & $C_x$, $A_z$   & 
$\begin{pmatrix} \alpha_{xx} &  & \\
                  & \alpha_{yy} & \\
                  & & \alpha_{zz} \\
\end{pmatrix}
$
& $(0,0,0)$   & $\varnothing$ \rule{0pt}{0.9cm}\\
$B_{1u}$ & Co, Fe & $C_y$          & 
$\begin{pmatrix}  &\alpha_{xy} & \\
                  \alpha_{yx}& & \\
                  & & \\
\end{pmatrix}
$
& $(0,0,T_z)$ & 0 \rule{0pt}{0.9cm}\\
$B_{2u}$ & Ni     & $C_z$, $A_x$   &  
$\begin{pmatrix}  & & \alpha_{xz} \\
                  & & \\
                  \alpha_{zx}& & \\
\end{pmatrix}
$
& $(0,T_y,0)$ & 0  \rule{0pt}{0.9cm}\\
 \hline
 \hline
\end{tabular}
\caption{Summary of the measured primary ($C$-type) magnetic ordering, and the resulting additional
magnetic orderings, toroidal and monopole moments, and components of the magnetolectric tensor (ME), 
obtained by symmetry analysis for the Li$M$PO$_4$ series.}
\label{Eric_Symmetry}
 \end{small}
  \end{center}
\end{table}

While it is at first sight tempting to describe LiCoPO$_4$ and LiFePO$_4$ as {\it non-monopolar}, this is not strictly correct, as we 
discuss next. 
First, we note that in the LiMPO$_4$ family, the P atom and the O$_1$ and O$_2$ atoms also occupy $4c$ sites, and so follow the
same symmetry transformations as the transition metal ions. This means that for LiMnPO$_4$ and LiNiPO$_4$ local monopoles are
allowed on these atoms. Of the remaining sites, the $4a$ of Li have only $\i$ as a symmetry operation, and the $8d$ sites of the O$_3$ 
have no site symmetry. In Table \ref{tab:LiMPO4-character} we also list the symmetries and possible monopole orderings of the 
$4a$ and $8d$ sites. We find that for the $A_{1u}$ irreducible representation of LiMnPO$_4$, the monopoles on Li and O$_3$ have the same ferromonopolar ordering as the Mn sites. 
Likewise, for LiNiPO$_4$, in which the Ni sites have antiferromonopolar ordering, an antiferromonopolar
ordering of the Li and O$_3$ monopoles is also found. Most notably, for LiFePO$_4$ and LiCoPO$_4$, which
have non-monopolar transition metal $4b$ sites, antiferromagnetically 
ordered monopoles are allowed on the $4a$ and $8d$ sites. 

In the next section we use first-principles density functional theory to 
calculate the magnitudes of these various contributions.

\subsection{Density functional calculations of atomic site monopoles and macroscopic monopolizations}

Our calculations were done using the local spin density approximation with an additional Hubbard $U$ correction on the 
transition metal sites (the LSDA$+U$ method). We took values of $U$=5eV and $J$=0.75eV for all systems; these values
correctly reproduce the experimentally reported magnetic orderings and anisotropies. For structural optimizations
we used the Vienna ab initio simulation package (VASP) \cite{Kresse/Furthmueller_PRB:1996} with a plane-wave
basis set and projector augmented wave\cite{Kresse/Joubert:1999} potentials. Our energy cutoff and $k$-point grid
were 500 eV and $2\times 2\times 4$ respectively. We used default VASP PAW potentials with the following electrons
in the valence: Li (1s, 2s), O (2s, 2p), P (3s, 3p), Co (3d, 4s), Mn, Fe and Ni (3p, 3d, 4s). Structural relaxations 
were performed in the absence of spin-orbit coupling.
For the monopole calculations we used the structures obtained form the VASP code, then used the linearized augmented 
plane wave (LAPW) method as implemented in the ELK code\cite{ELK} with spin-orbit coupling included to calculate
the charge and spin density. We used a basis set of $l_{max(apw)}=10$, a $9 \times 5 \times 5$ k-point sampling of 
the Brillouin zone and took the product of the muffin tin radius and the maximum reciprocal lattice vector to be 7.5. 
To calculate the atomic site monopoles ($a^{\text{as}}$) we decomposed the 
the density matrix into tensor moments as described in Section~\ref{Calculation_and_Measurement} 
\cite{Bultmark_et_al:2009} and evaluated the $d-p$ matrix elements for 
the transition metal atoms and the $p-s$ matrix elements for the Li, P and O atoms. 

In Table~\ref{Results} we report our calculated local atomic site monopoles $a^\text{as}$, for the series of transition metal phosphates,
as well as the local moment contribution, $a^\text{lm}$. Note that the orbital component makes no contribution by symmetry to the atomic site 
monopoles, and its magnitude is negligible in the local moment monopole of the ferromonopolar LiMnPO$_4$ because of the half-filled Mn$^{2+}$
$d$ shell. We also report the total macroscopic monopolizations, normalized to the unit volume, $A$. 

The first thing to note is that, in the ferromonopolar case of LiMnPO$_4$, the local moment monopole is as expected
considerably larger -- by around three orders of magnitude -- than the atomic site monopoles. The value of the local
moment monopole in one four-formula unit unit cell is 2.09 $\mu_B \AA$, whereas the local atomic site monopoles are 
all around 10$^{-3} \mu_B \AA$. Even when summed over all the atomic sites, the contribution from the atomic site
monopoles is still only $8.52 \times 10^{-3} \mu_B \AA$; it is so small in part because of cancellations between site monopoles of
different sign. The macroscopic monopolization, $A$, which is the total monopole per unit volume, then derives
almost entirely from the local moment contribution. We obtain a value of $A= 6.95 \times 10^{-3}  \mu_B/\AA^2$  modulo the monopolization
increment of $11.54 \times 10^{-3} \mu_B /\AA^2$ $\mu_B/\AA^2$. Note that, since we treat the magnetic moments as collinear along a lattice
vector there is just one monopolization increment.

\begin{figure}
\centerline{\includegraphics[width=0.95\columnwidth]{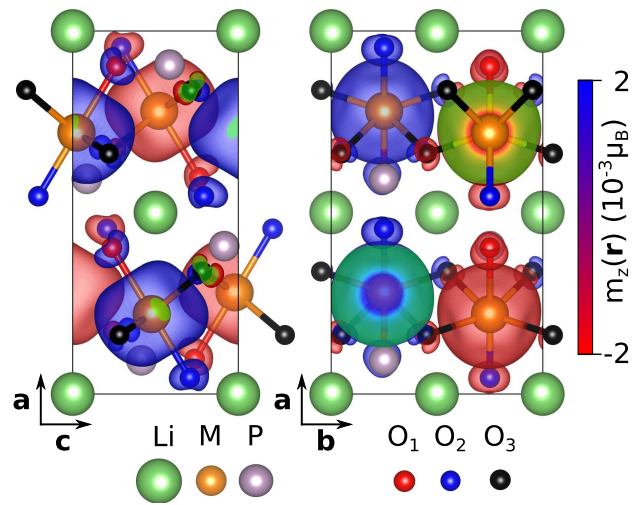}}
\caption{Calculated magnetization density isosurface for LiNiPO$_4$. The blue and
red surfaces correspond to positive or negative density, respectively. }
\label{mag_density}
\end{figure}

For the other compounds a net monopolization is forbidden by symmetry, and so the local moment monopole and the
total monopolization are both formally zero. We find, however, non-zero values for those atomic site monopoles 
that are allowed by symmetry, always with the appropriate symmetry-allowed antiferromagnetic ordering. 
Particularly interestingly, we find that when atomic site monopoles
are symmetry allowed on the P and O atoms, they are comparable to or larger than the values on the transition metals.
The relative sizes of the atomic site monopoles can be understood from inspection of the magnetization 
density: In Fig.~\ref{mag_density} we show the isosurface of our calculated magnetization 
density at 0.00125 $\mu_B / \AA\ ^3$ for
LiNiPO$_4$, with blue and red surfaces indicating positive and negative density, as well as a slice
through the magnetization density coinciding with the Ni site positions. The small deviation from a
perfectly spherical distribution around the Ni atom is indicative of the monopolar and other non-dipolar
multipolar contributions. It is clear that the magnetization density around the oxygen atoms, while
smaller in magnitude, is more non-spherical than that around Ni. In particular, the magnetization density
changes sign at the O$_3$ sites, indicating a highly non-spherical magnetization density which is 
consistent with their having the largest atomic site monopoles, 
The atomic site monopole on Li, although non-zero by symmetry for every case, is always small, consistent 
with the highly ionic nature of the Li$^+$ ion; since the charge density around the Li ions is close
to zero, the magnetization density is too (Fig.~\ref{mag_density}).
Finally we note that the atomic site monopole on Ni in LiNiPO$_4$ is one order of magnitude smaller than that on Mn in 
LiMnPO$_4$, even though its local magnetic dipole moments is only $\sim$2.5 times smaller. Our initial computer
experiments suggest that this is partly a result of the different magnetic anisotropy in the two cases, as a calculation
with the Ni moments constrained to have the same orientation as those of Mn in LiMnPO$_4$ yields increased atomic site
monopoles. A detailed study of the factors that determine the magnitudes of atomic site monopoles will be the subject
of future work.

\begin{table}[htbp!]
\centering
\begin{ruledtabular}
\begin{tabular}{lrrrr}
  & Mn   & Fe  & Co  & Ni  \rule[-1ex]{0pt}{3.5ex}  \\
$a^{\text{as}}$ ($\times 10^{-3} \mu_B \AA$)  &    &   &  &    \\
\hline
M        &  1.94  & 0.00 & 0.00 & 0.09  \rule[-1ex]{0pt}{3.5ex} \\
Li       &  0.06  & 0.03 & 0.04 & 0.01  \rule[-1ex]{0pt}{3.5ex} \\
P        &  3.20  & 0.00 & 0.00 & 0.49  \rule[-1ex]{0pt}{3.5ex} \\
O$_1$    & -7.68 & 0.00 & 0.00 & -3.14  \rule[-1ex]{0pt}{3.5ex} \\
O$_2$    &  7.14 & 0.00 & 0.00 & 4.10 \rule[-1ex]{0pt}{3.5ex} \\
O$_3$    & -1.26 & -6.02& -6.74 &-7.63 \rule[-1ex]{0pt}{3.5ex} \\
\hline
$\sum a^{as}$ ($\times 10^{-3} \mu_B \AA$)& 8.52 & 0.00 & 0.00 & 0.00 \rule[-1ex]{0pt}{3.5ex} \\
$a^{lm}$ ($\times 10^{-3} \mu_B \AA$)& 2091.94 & 0.00 & 0.00 & 0.00 \rule[-1ex]{0pt}{3.5ex} \\
\hline
$A^{\text{lm}}$ ($\times 10^{-3}\mu_B$/\AA$^2$) & 6.92  & 0.00 & 0.00 & 0.00  \rule[-1ex]{0pt}{3.5ex}\\
$A$ ($\times 10^{-3}\mu_B$/\AA$^2$)   & 5.95 & 0.00 &0.00 & 0.00  \rule[-1ex]{0pt}{3.5ex}\\
\end{tabular}
\caption{Calculated atomic site monopoles, local moment monopoles, and macroscopic monopolizations for the Li transition metal phosphates.
}
\label{Results}
\end{ruledtabular}
\end{table}

\section{Multiferroic Free Energy with monopole contributions}
\label{GLTheory}

As stated above,
from a macroscopic symmetry point of view, the symmetries which allow
for a macroscopic monopolization are identical with that allowing for
a diagonal component of the linear magnetoelectric effect tensor. 
In this section, we develop the relationship between these two quantities 
by analyzing the following free energy expression:
\begin{multline}
\label{free-energy}
U = \frac{1}{2 \epsilon} P^2 - \vec{P}\cdot\vec{E} + \frac{1}{2 \chi} M^2 -
\vec{M}\cdot\vec{H} \\
+ \frac{1}{2} \beta A^2 + \frac{1}{4}\gamma A^4 + c A \vec{P} \cdot \vec{M}
\quad ,
\end{multline}
where $\epsilon$ and $\chi$ are the electric and
magnetic susceptibilities, $\beta$ and $\gamma$ are temperature-dependent
coefficients, and $c$ determines the strength of the magnetoelectric
coupling. 
This is the simplest possible free energy expression that can
simultaneously describe (i) a phase transition from a para-monopolar
($A=0$) into a ferromonopolar phase ($A \neq 0$), (ii) the
coupling of the electric polarization $\vec{P}$ and the magnetization
$\vec{M}$ to the electric field $\vec{E}$ and the magnetic field
$\vec{H}$, respectively, and (iii) a coupling between the electric
polarization, the magnetization, and the monopolization.
Note that only the magnetization and the polarization
couple to $\vec{H}$ and $\vec{E}$, the monopolization in general does
not couple to any homogeneous external fields, in agreement with the
fundamental definitions discussed in Sec.~\ref{MultipoleExpansion}. 
The trilinear form of the coupling term in
Eq.~(\ref{free-energy}) is the lowest possible order that is
compatible with the overall space and time reversal symmetries. 
Since our purpose here is to discuss the new features arising from this trilinear
coupling, we leave for future work the analysis of gradient terms in the free energy 
that would be required to 
describe for example variations in monopolization, magnetization or polarization 
at domain walls. The
equilibrium values for $\vec{P}$ and $\vec{M}$ can be obtained by
minimizing Eq.~(\ref{free-energy}). This leads to:
\begin{equation}
\label{pol}
\vec{P} = \epsilon ( \vec{E} - c A  \vec{M} )
\end{equation}
and
\begin{equation}
\label{mag}
\vec{M} = \chi ( \vec{H} - c A \vec{P} ) \quad .
\end{equation}
If one inserts Eq.~(\ref{mag}) into Eq.~(\ref{pol}) one obtains (to
leading order in $A$):
\begin{equation}
\label{PME}
\vec{P} = \epsilon ( \vec{E}  - \chi c A \vec{H} ) \quad .
\end{equation}
The last term in Eq.~(\ref{PME}) is a symmetric linear
magnetoelectric effect proportional to the monopolization. Thus, the
presence of the trilinear coupling term between monopolization,
magnetization, and polarization in Eq.~(\ref{free-energy}) gives rise
to a diagonal magnetoelectric effect $\vec{P} = \alpha \vec{H}$
in the ferromonopolar phase, with
\begin{equation}
\alpha_{ii} = \alpha_{jj} = \alpha_{kk} = \epsilon \chi c A \quad .
\end{equation}
(Note that an off-diagonal magnetoelectric effect is obtained from
a trilinear coupling between toroidization, magnetization and polarization,
as discussed in Ref.~\onlinecite{Ederer/Spaldin:2007}). 

Conversely, the presence of a monopolar contribution can be inferred from the existence
of a diagonal linear magnetoelectric response, the magnitude of which is determined by
the product of the dielectric susceptibility, magnetic permeability, monopolization
and the strength of the coupling between $A$, $\vec{P}$ and $\vec{M}$.
If the linear magnetoelectric response is diagonal and isotropic, then
there can be no quadrupolar contributions and the response arises entirely from monopolar
contributions.
We see also from Eqn.~\ref{PME} that in the case of antiferromonopolar ordering,
a homogeneous magnetic field will induce a finite-$q$ polarization. Such a relationship
could be used in the case of $q=\pi/a$, to provide a more fundamental definition of an
antiferroelectric in simultaneously antiferromonopolar systems, than the current unsatisfactory working 
definition based on the observation of double-loop hysteresis.
Finally we mention that an additional interesting consequence of the relationship between the monopolization
and the diagonal magnetoelectric effect
is the induction of monopoles by electric charge. This has been discussed previously in the
context of axion electrodynamics\cite{Wilczek:1987}, and is currently being revisited in the
context of topological insulators\cite{Essin/Moore/Vanderbilt:2009}.

\section{Summary, conclusions, and outlook}
\label{conclusions}

In summary we have presented a theoretical analysis of magnetoelectric monopoles
in bulk periodic solids. We introduced the term ``monopolization'' to describe
the monopole moment per unit volume, and considered two contributions, one arising from the 
local variation in magnetization density around the atom and the second from the 
distribution of localized magnetic dipole moments throughout the unit cell. We 
found that the latter dominates the total monopolization in transition metal compounds 
with ferromonopolar ordering.
We showed that, for ferromonopolar materials, periodic boundary conditions lead to a
multivaluedness of the monopolization, suggesting that only differences in
monopolization are well-defined observable macroscopic quantities.
We found also that care must be taken in evaluating such monopolization 
differences: For example in the 
example of the distorted one-dimensional antiferromagnetic chain
discussed in Sec.\ref{chain}, the change in monopolization due to
a structural distortion can be calculated straightforwardly, whereas
in the case of a magnetic moment reversal one has to subtract the
improper monopolization change that is caused by the corresponding
change in the monopolization increment. 

Quantitative measurements of monopolizations are challenging.
The atomic site monopolization can in principle be detected using resonant
x-ray scattering, although the experimental constraints are rather rigorous
and a suitable material for such an experiment has not yet been identified. 
In particular, for most space group symmetries the sites that allow an
atomic site monopole also allow an atomic site quadrupolar component, and disentangling
the two contributions is not straightforward \cite{Staub_et_al:2009}.
This problem can be circumvented by selecting materials with an isotropic diagonal
magnetoelectric response \cite{Hehl_et_al:2008}, however few such materials have
been identified to date.
Even more problematic is the question of how to measure the macroscopic 
local moment monopolization. According to the fundamental definition of the
monopole moment, this is in principle possible by measuring the effect 
on a sample of a diverging magnetic field, however such a field is not accessible. 
It is possible that earlier observations of a quadrupolar magnetic field around a 
spherical sample of the prototypical diagonal magnetoelectric Cr$_2$O$_3$ 
\cite{Astrov/Ermakov:1994,Astrov_et_al:1996}
also incorporate a monopolar contribution; the theory underlying these measurements
will be revisited in future work \cite{Dzyaloshinskii:1992}.
It has also been recently proposed that signatures of monopolar behavior will
manifest in the transport properties of diagonal magnetoelectrics \cite{Khomskii:2013}

An open question, both for ferrotoroidic and ferromonopolar materials is whether the 
toroidal moment or monopole moment respectively can be a primary order parameter,
or is always secondary to an antiferromagnetic or structural ordering.
Currently no case has been identified even theoretically in which the monopolization 
is non-zero while there is no magnetic ordering, although it is possible that some ``hidden-order
parameter'' materials that are of current interest might prove to fall into this
class \cite{Shekhter/Varma:2009}.
The fact that the monopole order parameter is a scalar might be helpful in
distinguishing responses that arise from the antiferromagnetism from those of
the monopole, in cases where the antiferromagnetic order parameter is a vector.
Within the class of secondary ferromonopolar materials, it is also an open question
whether there is a fundamental difference between the case in which the primary 
order parameter is the AFM ordering, and that where it is a structural
phase transition from a centrosymmetric antiferromagnet (which does not allow
monopolization) to a non-centrosymmetric monopolar state. 

Finally, we mention that it has been argued that ferrotoroidicity is a key concept for fitting
all forms of ferroic order in a simple fundamental scheme based on the
different transformation properties of the corresponding order
parameters with respect to time and space inversion (see
Refs.~\onlinecite{Schmid:2003,Schmid:2004,VanAken_et_al:2007}, in
particular Fig.~2 in Ref.~\onlinecite{VanAken_et_al:2007}). It is clear
that from a symmetry point of view, that the monopolization could play
a similar role, since a ferromonopolar material also breaks both space-inversion and
time-reversal symmetry.  As a result the nonlinear optical techniques
used in Ref.~\onlinecite{VanAken_et_al:2007} to identify ferrotoroidic
ordering are sensitive also to the monopolar symmetry breaking, and could provide 
indirect evidence for the presence of monopolization. 
In addition, the four fundamental forms of ferroic order,
with order parameters transforming according to the four different representations
of the ``parity group'' generated by the two operations of time and space
reversal\cite{Ascher:1974} could be chosen to be ferroelasticity, ferroelectricity, ferromagnetism, and
ferromonopolicity (rather than ferrotoroidicity). 
Whether the scalar nature of the monopole, compared with the vector nature of the toroidal moment,
makes this choice more or less appropriate is an open question.

\begin{acknowledgments}
This work was supported financially by the ETH Z\"urich (NAS, MF and EB), by the ERC Advanced Grant 
program, No. 291151 (NAS and EB), by the Max R\"ossler Prize of the ETH Z\"urich (NAS), 
Nordita (AB), US DoE (AB) and  the Swedish Research Council (AB and LN). EB is a Research Associate of the
Fonds de la Recherche Scientifique, FNRS, Belgium.
NAS thanks Nordita, the Nordic Institute for Theoretical Physics, for their hospitality 
during a visit where much of this work was performed. 
\end{acknowledgments}

\end{document}